\documentclass[twocolumn,amsmath,amssymb,nofootinbib,aps,pre,unsortedaddress]{revtex4-2}

\usepackage{graphicx,color}
\definecolor{brown}{rgb}{0.63,0.27,0.18}
\definecolor{orange}{rgb}{1.00,0.65,0.00}
\usepackage{afterpage} %<--- Help fixing the float problems due to the *many* figures at the end; place the command *\afterpage{\clearpage}* in front of some *\begin{figure*}[p!]* at the end; see: https://www.guitex.org/home/en/forum/5-tex-e-latex/83467-problema-too-many-unprocessed-floats

\usepackage{moresize}
\usepackage{dcolumn}
\usepackage{bm,multirow}
\makeatletter
\newcommand*{\balancecolsandclearpage}{%
  \close@column@grid
  \twocolumngrid
}
\makeatother
\usepackage{color}

\begin{document}

\title{Universal time and length scales of polar active polymer melts} %: Introducing a dynamic lattice model}
%

%%%
\author{Mattia Alberto Ubertini}
\email{mubertin@sissa.it}
\email{Current affiliation: Friedrich Miescher Institute for Biomedical Research (FMI), Basel, Switzerland}
\affiliation{Scuola Internazionale Superiore di Studi Avanzati (SISSA), Via Bonomea 265, 34136 Trieste, Italy}

\author{Emanuele Locatelli}
\email{emanuele.locatelli@unipd.it}
\affiliation{Department of Physics and Astronomy, University of Padova, Via Marzolo 8, I-35131 Padova, Italy and INFN, Sezione di Padova, Via Marzolo 8, I-35131 Padova, Italy}

\author{Angelo Rosa}
\email{anrosa@sissa.it}
\affiliation{Scuola Internazionale Superiore di Studi Avanzati (SISSA), Via Bonomea 265, 34136 Trieste, Italy}
%%%

%

\begin{abstract}
We present an in-depth multi-scale analysis of the conformations and dynamics of polar active polymers, comparing very dilute and very dense conditions.
We unveil characteristic length and time scales, common to both dilute and dense systems, that recapitulate the conformational and dynamical properties of these active polymers upon varying both the polymer size and the strength of the activity.
Specifically, we find that a correlation (or {\it looping}) length characterises the polymer conformations and the monomer dynamics.
Instead, the dynamics of the center of mass can be fully characterised by the end-to-end mean-square distance and by the associated relaxation time.
As such, we show that the dynamics in melts of polar active polymers are not controlled by entanglements but only by the strength of the self-propulsion. 
\end{abstract}
%

%%%
\maketitle

%\section{Introduction}\label{sec:Intro-Active}
{\it Active matter} consists of systems whose fundamental units are able to transduce energy into persistent movement leading to features very different from the passive counterparts~\cite{ramaswamy2010mechanics,marchetti2013hydrodynamics}.
Examples of such systems abound in nature, from the macroscopic scale, such as bird flocks or fish shoals~\cite{vicsek2012collective}, down to the sub-cellular level, the main example being the cytoskeleton~\cite{fletcher2010cell}. 
Of late, significant attention has focused on a specific subset of active matter -- {\it active polymers}~\cite{winkler2020physics}.
This interest arises from their biological relevance, as many biological systems feature molecular motors, which serve as prototypical examples of active matter by converting energy from chemical reactions into motion along biological filaments, such as DNA or RNA.
For instance, DNA polymerase traverses DNA strands during replication, while ribosomes actively synthesize proteins by sliding along RNA filaments~\cite{alberts2002cytoskeleton}.
Experimentally, anomalous diffusion of chromatin loci has been observed~\cite{weber2012nonthermal} and showed to be caused by non-equilibrium processes related to enzymatic activity.
A further example at the micrometric scale is given by cilia and flagella, used in mono- as well as multi-cellular organisms for transport and locomotion~\cite{gilpin2020multiscale}.
Furthermore, active polymer models have been used to rationalise the individual and collective properties of filamentous bacteria and parasites~\cite{patra2022collective,faluweki2023active,kurjahn2024collective}. 
On the other hand, active filamentous systems have also attracted attention at the macro-scale, as worms~\cite{deblais2023worm} seem to provide an interesting experimental system with anomalous emerging properties~\cite{deblais2020phase,deblais2020rheology,ozkan2021collective,patil2023ultrafast}.
In all cases, what makes active filaments interesting is that the energy input at the monomer level changes the typical conformation of the whole filament and influences its dynamics in a multi-scale fashion.
This is true for a single filament; however, at finite density, it reverberates on the organisation of the entire system~\cite{Smrek2020,patra2022collective,dunajova2023chiral,miranda2023self}.
Consequently, understanding the multi-scale organisation of active polymer systems represents an important challenge.

In this paper, we focus on polar (or tangential) active polymers, where a self-propulsion force is applied to each monomer parallely to the local backbone tangent.
In this case, significant progress has been achieved in characterizing dilute active linear and ring polymers~\cite{isele2015self,bianco2018globulelike,locatelli2021activity,foglino2019non,philipps2022tangentially,philipps2022dynamics,vatin2024conformation}.
Quite surprisingly, less work has been devoted to dense, entangled polymer systems (melts), feature that is relevant in many contexts, as for instance in chromatin organization~\cite{RosaEveraersPlos2008}.
In particular, the efforts were restricted to the limit of {\it small} activity~\cite{tejedor2019reptation,tejedor2020dynamics,tejedor2023molecular}, with the recent exception of Ref.~\cite{Breoni2023} exploring the consequences of large activity for linear viscoelasticity.
Here, we investigate how activity impacts on the size, shape and dynamics of entangled polymer chains in melt and compare them {\it vis-\`a-vis} with their counterparts in dilute conditions, to highlight the main differences and the interplay between entanglements and activity.

Specifically, we employ Langevin molecular dynamics computer simulations (for details, see Section~\ref{sec:PolymerModelMethods} in Supporting Information (SI)) of a well established model for polar active polymers~\cite{bianco2018globulelike} in the presence of both, Brownian thermal forces, mimicking the effect of the solvent, {\it and} active local tangential forces~\cite{bianco2018globulelike} of constant norm $f_a$ imparted on every monomer, except the first and the last one, of each chain in the system.
In order to express the relative strength of the activity against the thermal fluctuations, we introduce the usual {\it a}dimensional P{\'e}clet number ${\rm Pe} \equiv {f_a \sigma}/{(\kappa_B T)}$ where $T$ is the temperature of the environment and $\kappa_B$ is the Boltzmann constant.
Here we choose ${\rm Pe} = 1, 5, 10, 20$, {\it i.e.} from low to high activity regime and, for reference, we compare these situations to the purely {\it passive} case ${\rm Pe}=0$.
We consider systems of hundreds ($250$, at least) polymer chains of, respectively, $N=100, 200, 400, 800$ monomers per chain at the overall monomer density $\rho = 0.85\sigma^{-3}$ ($\sigma$ is the monomer diameter) corresponding to typical melt conditions~\cite{KremerGrest-JCP1990}.
For comparison, we simulate the same chains at very dilute conditions. 
A more detailed account of the polymer systems considered in this work (including their preparation and proper equilibration) is provided in Sec.~\ref{sec:PolymerModelMethods} in SI.

We characterize the folding of polymer chains across all scales employing the mean-square internal distance for monomer pairs with contour length separation $n$,
\begin{equation}\label{eq:R2n}
\langle R^2(n) \rangle \equiv \frac1{N-n} \sum_{n'=1}^{N-n} \langle \left( {\vec r}_{n'+n} - {\vec r}_{n'} \right)^2 \rangle \, , 
\end{equation}
where ${\vec r}_n$ ($n=1, ..., N$) is the spatial position of monomer $n$ and the brackets $\langle \cdot \rangle$ denote the ensemble average; we further consider the corresponding (local) scaling exponent,
\begin{equation}\label{eq:DefineNu}
\nu(n) \equiv \frac12 \, \frac{\log(\langle R^2(n+1) \rangle) - \log(\langle R^2(n-1) \rangle)}{\log(n+1) - \log(n-1)} \, .
\end{equation}
For passive systems ({\it i.e.}, ${\rm Pe}=0$), $\langle R^2(n) \rangle$ display monotonous behavior (see corresponding panels in Fig.~\ref{fig:R2-vs-n} in SI) and the scaling exponent, although within the limits of finite-size effects, agree with the established~\cite{RubinsteinColbyBook} asymptotic values $\nu_{\rm dilute} \simeq 0.6$ and $\nu_{\rm melt} = 0.5$.
Conversely, the behavior of $\langle R^2(n) \rangle$ for active systems appears more complex (see corresponding panels in Fig.~\ref{fig:R2-vs-n} in SI). 
At fixed ${\rm Pe}$ and $N$, we report three regimes for small, intermediate and large $n$, with asymptotic $\nu$ (see insets) {\it smaller} than the passive chains (we will expand on this point later).
However, and most remarkably, for the same ${\rm Pe}$, $\langle R^2(n) \rangle$ depends, at intermediate contour separations, also on the {\it total} chain length $N$, a feature completely absent in passive systems.

In order to rationalize this peculiar behavior, we look at the bond-vector correlation function as a function of the contour length separation $n$~\cite{foglino2019non,locatelli2021activity,fazelzadeh2023effects},
\begin{equation}\label{eq:tgtgCorrel}
c(n) \equiv \frac1{\langle {\vec t}^{\, 2} \rangle} \frac{\sum_{n'=1}^{N-1-n} \, \langle {\vec t}_{n'+n} \cdot {\vec t}_{n'} \rangle}{N-1-n} \, ,
\end{equation}
where ${\vec t}_n \equiv {\vec r}_{n+1} - {\vec r}_n$ ($n=1, ..., N-1$) is the oriented bond-vector and $\langle {\vec t}^{\, 2} \rangle = \sum_{n'=1}^{N-1} \, \langle {\vec t}_{n'}^{\, 2} \rangle / (N-1)$ is the mean-square bond length, {\it i.e.} $c(0)=1$ by construction. Since $c(n)$ and $\langle R^2(n) \rangle$ are related to each other via the relation $c(n) = \frac12 \frac{{\rm d}^2}{{\rm d}n^2} \langle R^2(n) \rangle$, features of $\langle c(n) \rangle$ may help understanding the observed phenomenon.
Indeed, in passive systems (see top row in Fig.~\ref{fig:BCF-vs-n} in SI), $c(n)$ displays a power-law decay for both dilute and melt conditions, with characteristic exponents ($c(n)\sim n^{-0.824}$ for dilute systems and $c(n)\sim n^{-3/2}$ for melts) in good agreement with previous results~\cite{WittmerPRL2004}. 
On the contrary, the distinct scaling regimes displayed by $\langle R^2(n) \rangle$ for ${\rm Pe} >0$ are mirrored by the non-monotonous behavior of $c(n)$ (second to last row in Fig.~\ref{fig:BCF-vs-n} in SI).
Indeed, for ${\rm Pe} >0$, $c(n)$ exhibits a distinct {\it negative} minimum, {\it i.e.} bond-vectors become anti-correlated, at some characteristic length scale $n \equiv n_{\rm min}$.
Afterwards, correlations decay back to zero, yet with a much steeper behavior ($c(n) \sim n^{-3}$, roughly) than in passive conditions. 

\begin{figure}
\includegraphics[width=0.50\textwidth]{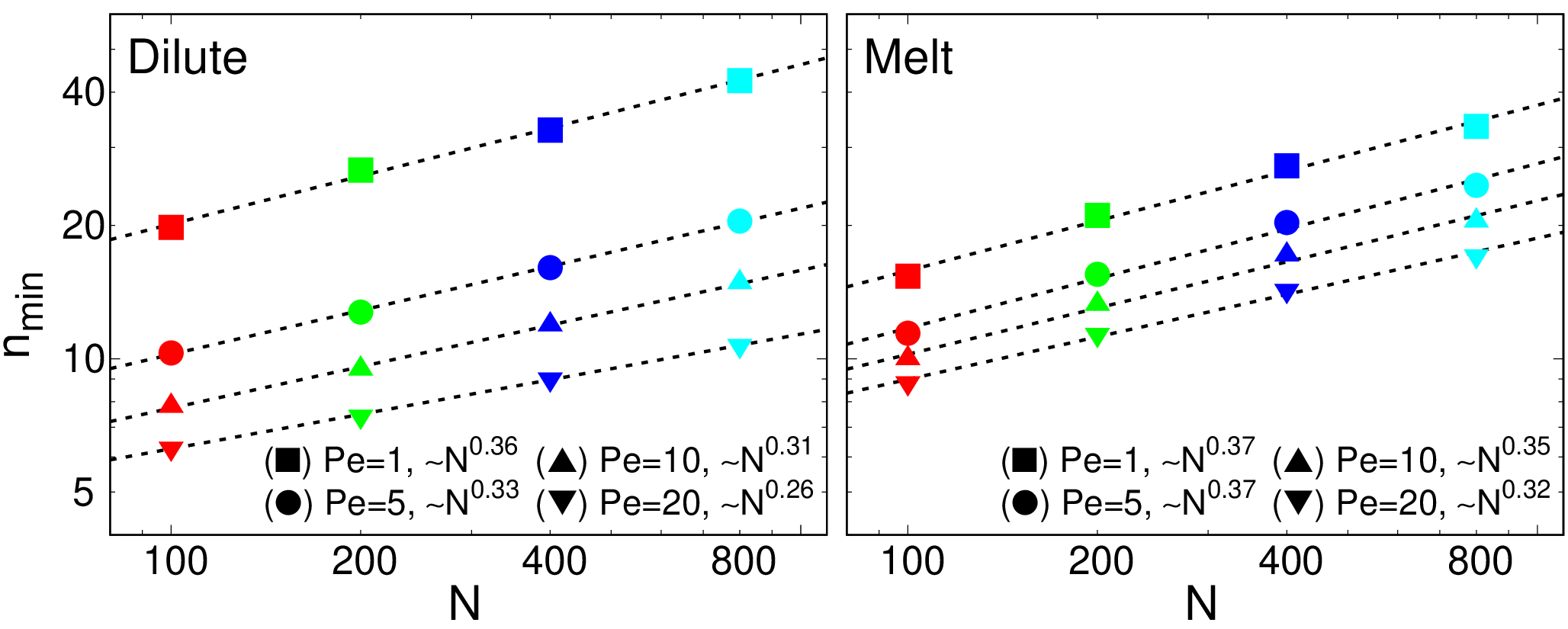}
\caption{
Contour distance $n_{\rm min}$ (symbols), corresponding to the contour length position of the minimum value of the bond-correlation function $c(n)$ (Eq.~\eqref{eq:tgtgCorrel}) for P\'eclet number ${\rm Pe}>0$, as a function of the polymer total length $N$.
Each dashed line corresponds to the best fit of the relative data to the power-law~\eqref{eq:NminPowerLaw}, whose values for the exponent $\alpha$ are reported in the legend (and in Table~\ref{tab:nMinFitParams} in SI).
}
\label{fig:nminVSN}
\end{figure}
\begin{figure}%[!b]
\includegraphics[width=0.50\textwidth]{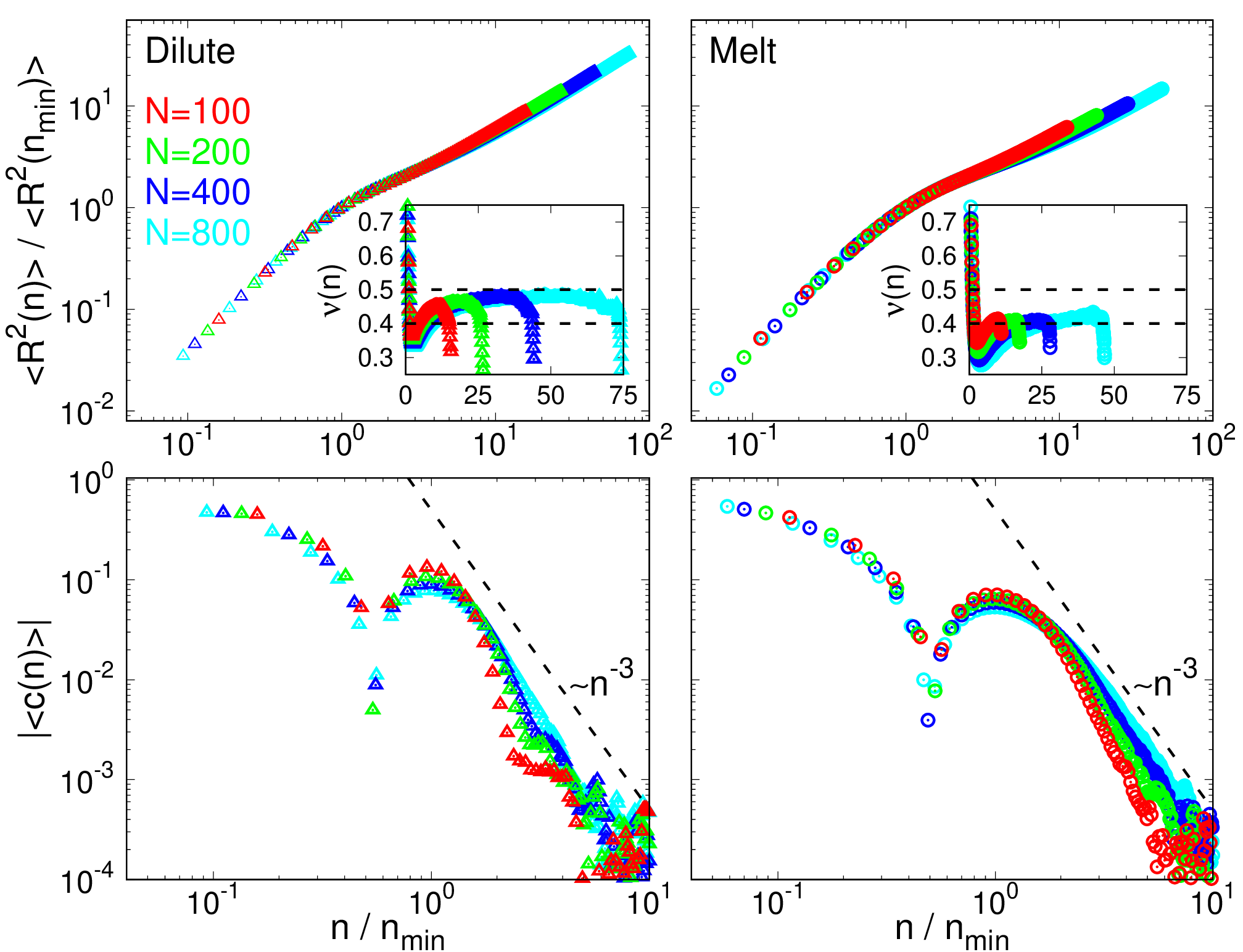}
\caption{
(Top, main)
Normalized mean-square internal distance, $\langle R^2(n) \rangle / \langle R^2(n_{\rm min}) \rangle$, as a function of normalized contour length separation, $n / n_{\rm min}$.
(Top, insets)
Local scaling exponent $\nu(n)$ (Eq.~\eqref{eq:DefineNu}) as a function of $n/n_{\rm min}$.
The dashed lines set the reference values $\nu=0.4$ and $\nu=0.5$.
(Bottom)
Bond-vector correlation function $\langle c(n) \rangle$  as a function of normalized contour length separation $n / n_{\rm min}$.
The asymptotic $\sim n^{-3}$-decay is a guide to the eye and purely indicative.
For all panels, ${\rm Pe}=20$ is fixed; symbols refer to dilute ($\triangle$) and melt ($\circ$) conditions, respectively.
}
\label{fig:R2-BCF-vs-n-scaling-Pe20}
\end{figure}

The presence of the negative minimum, entailing a length scale $n_{\rm min}$, is particularly suggestive as it implies~\cite{MullerWittmerCates2000,RosaEveraersPRL2014} that, both in dilute and melt conditions, active polymers tend to (double)fold into {\it looped} conformations of loop mean contour length approximately equal to $n_{\rm min}$. 
We thus report such emerging length scale in Fig.~\ref{fig:nminVSN}. We notice that, in both dilute and melt conditions, $n_{\rm min}$ {\it decreases} upon increasing $\mathrm{Pe}$ at fixed $N$. Conversely, at fixed $\rm Pe$, $n_{\rm min}$ grows with $N$ as a power-law.
We quantify this behavior by fitting our data (see dashed lines in Fig.~\ref{fig:nminVSN}) to the phenomenological power-law
\begin{equation}\label{eq:NminPowerLaw}
n_{\rm min} = n_0 \left( \frac{N}{n_0} \right)^\alpha \, ,
\end{equation}
where $n_0$ is a microscopic contour length scale that depends on $\rm Pe$ and where $0.259 \leq \alpha \leq 0.36$ for dilute chains and $0.32 \leq \alpha \leq 0.37$ for melts (see Table~\ref{tab:nMinFitParams} in SI for the values of the fit parameters).
More importantly, we claim that $n_{\rm min}$ is the {\it fundamental} length scale of active systems.
In fact, by normalizing the mean-square internal distances $\langle R^2(n) \rangle$ 
in terms of $\langle R^2(n=n_{\rm min}) \rangle$ and after rescaling $n$ by $n_{\rm min}$, the distinct sets of data for each $\rm Pe$ collapse onto one single master curve; the same happens with the bond-correlation functions $c(n)$ (see Fig.~\ref{fig:R2-BCF-vs-n-scaling-Pe20} and Fig.~\ref{fig:R2-BCF-vs-n-scaling-Pe>0} in SI).
After rescaling, $\langle R^2(n) \rangle$ and the associated exponent $\nu(n)$ (Eq.~\eqref{eq:DefineNu}) reveal the following noteworthy features:
(i) for $n / n_{\rm min} \lesssim 1$ both dilute and melt polymers are stiffened by the activity, {\it i.e.} $\nu(n) > 0.5$ (see insets in the panels for rescaled $\langle R^2(n) \rangle$ in Fig.~\ref{fig:R2-BCF-vs-n-scaling-Pe20} and Fig.~\ref{fig:R2-BCF-vs-n-scaling-Pe>0} in SI). 
Interestingly, by fitting $\langle R^2(n_{\rm min}) \rangle$ with a power-law $R_0^2 (N/n_0)^{\beta}$ and using the results of Eq.~\eqref{eq:NminPowerLaw}, one can derive the power-law exponent $\beta / \alpha$ for $\langle R^2(n_{\rm min}) \rangle$ as a function of $n_{\mathrm{min}}$, resulting in $\beta / \alpha >1$ systematically (Table~\ref{tab:nMinFitParams} in SI); 
(ii) for $n / n_{\rm min} \simeq 1$, the activity-induced looping of the polymer results in an unusually low value for $\nu$ as the polymer tends to become locally more compact; 
(iii) finally, in the long-chain limit ($n / n_{\rm min} \gg 1$) dilute and melt polymers exhibit the asymptotic values for $\nu(n)$, respectively $\nu(\infty) \simeq 0.5$ and $\nu(\infty) \simeq 0.4$. 
Quite interestingly, these asymptotic values appear to be independent of the activity level already for relatively moderate activities ${\rm Pe} \geq 5$.

\begin{figure}%[!b]
\includegraphics[width=0.50\textwidth]{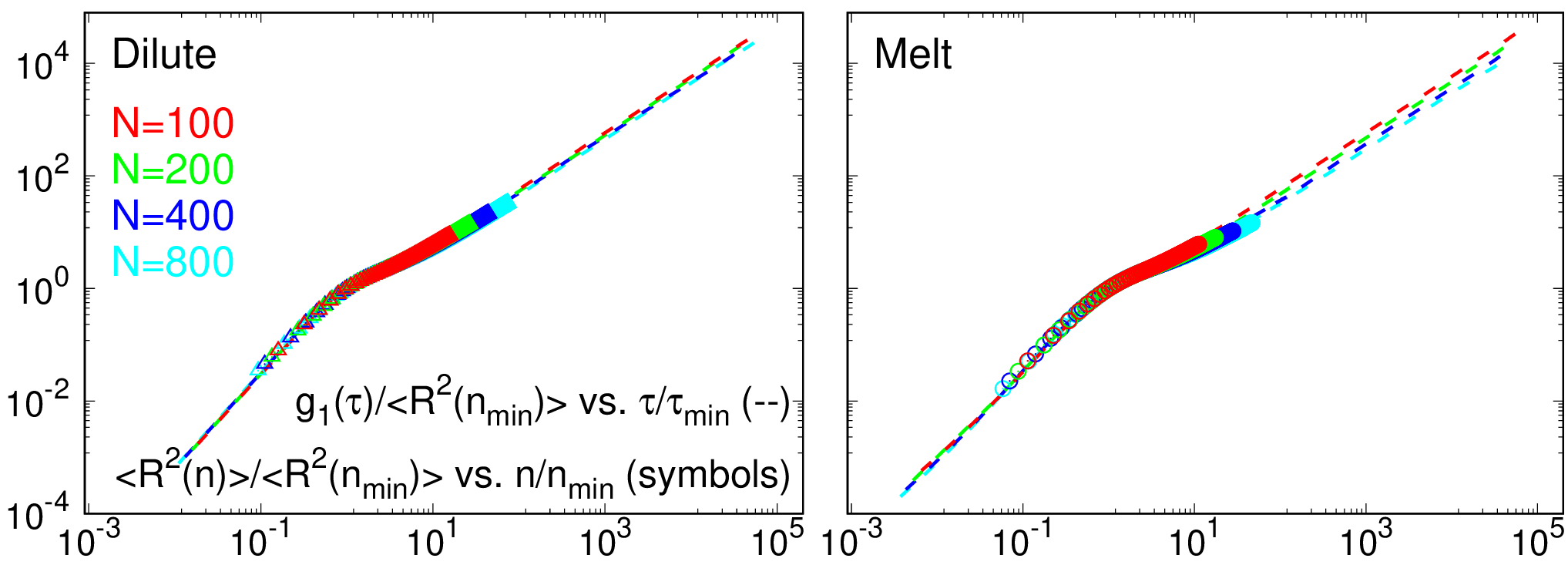}
\caption{
Monomer mean-square displacement, $g_1$, normalized by $\langle R^2(n_{\rm min})\rangle$ as a function of rescaled time $t/\tau_{\rm min}$ (lines) in comparison to normalized mean-square internal distance, $\langle R^2(n) \rangle / \langle R^2(n_{\rm min}) \rangle$, as a function of normalized contour length separation, $n / n_{\rm min}$ (symbols).
Symbols are as in Fig.~\ref{fig:R2-BCF-vs-n-scaling-Pe20} and ${\rm Pe}=20$.
}
\label{fig:g1-scaling-Pe20}
\end{figure}

We further show that $n_{\rm min}$ plays an important role in the characterization of the chain dynamics.
We consider the monomer mean-square displacement~\cite{KremerGrest-JCP1990},
\begin{equation}\label{eq:g1}
g_1({\tau}) \equiv \frac1N \sum_{n=1}^N \langle \left( {\vec r}_n(t+\tau) - {\vec r}_n(t) \right)^2 \rangle \, ,
\end{equation}
as a function of time $\tau$; we define the time scale $\tau_{\rm min}$ as $g_1(\tau_{\rm min}) = \langle R^2(n_{\rm min})\rangle$.
By expressing $g_1$ and the time $\tau$ in terms of, respectively, $\langle R^2(n_{\rm min})\rangle$ and $\tau_{\rm min}$, the different data superimpose perfectly on the normalized data for $\langle R^2(n) \rangle$ (Fig.~\ref{fig:g1-scaling-Pe20} for ${\rm Pe}=20$ and Fig.~\ref{fig:g1-scaling-Pe>0} in SI), namely monomers, on average, move {\it along} the contour length of their chain.
Overall, activity triggers a kind of motion where each monomer is, on average, pulled along the polymer local shape before the chain reorganizes into a new conformation.

\begin{figure}%[!t]
\includegraphics[width=0.50\textwidth]{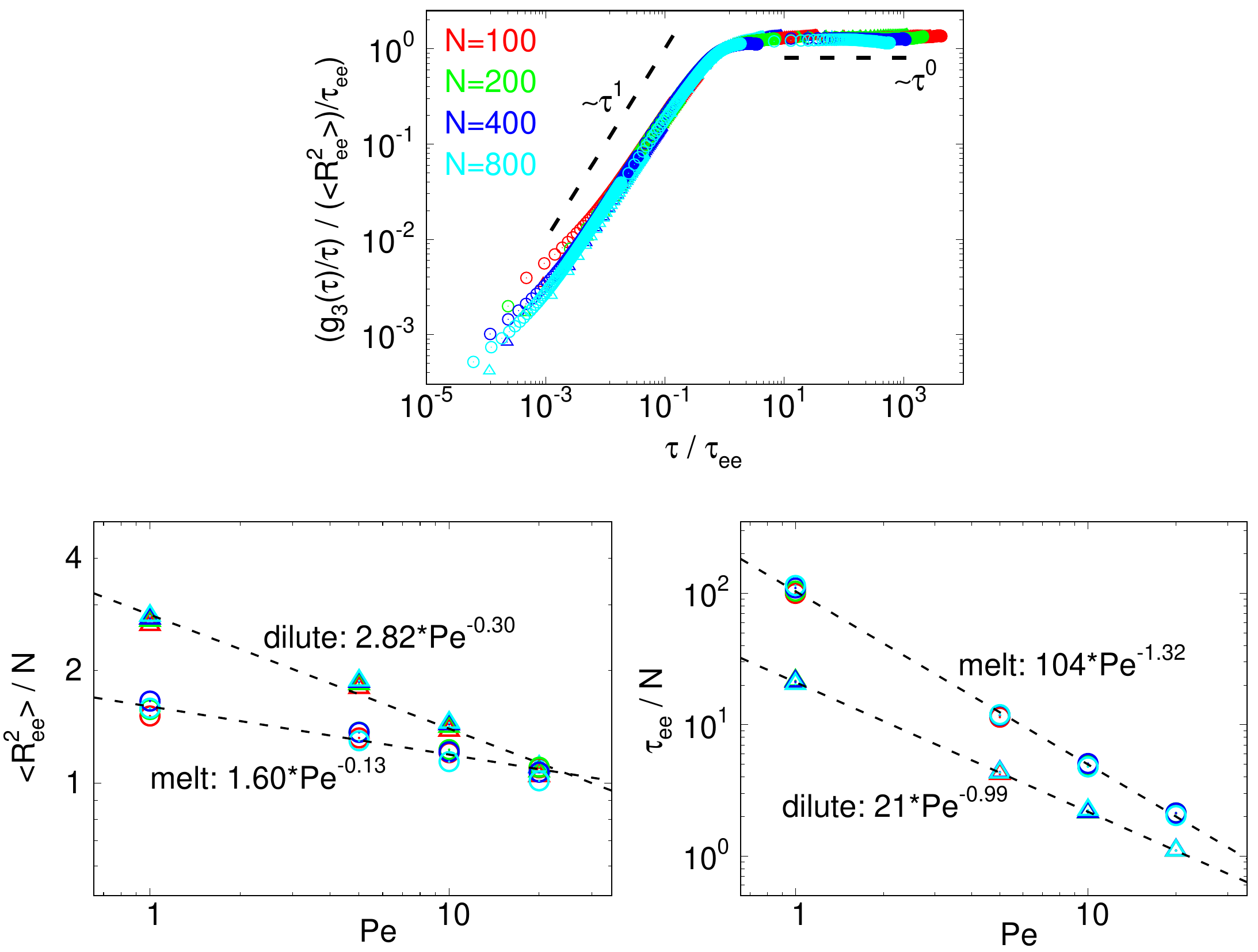}
\caption{
(Top)
Mean-square displacement of the chain centre of mass per unit time, $g_3(\tau) / \tau$, normalized by $\langle R_{\rm ee}^2 \rangle / \tau_{\rm ee}$ as a function of normalized time $\tau / \tau_{\rm ee}$ at fixed ${\rm Pe}=20$. 
(Bottom, left)
Chain mean-square end-to-end distance $\langle R_{\rm ee}^2\rangle$ normalized by $N$ as a function of $\rm Pe$ (symbols) and corresponding power-law best fits (lines).
(Bottom, right)
Chain relaxation time $\tau_{\rm ee}$ normalized by $N$ as a function of $\rm Pe$ (symbols) and corresponding power-law best fits (lines).
Symbols are as in Fig.~\ref{fig:R2-BCF-vs-n-scaling-Pe20}.
}
\label{fig:Dynamics-Active}
\end{figure}
\begin{figure}
\includegraphics[width=0.95\columnwidth]{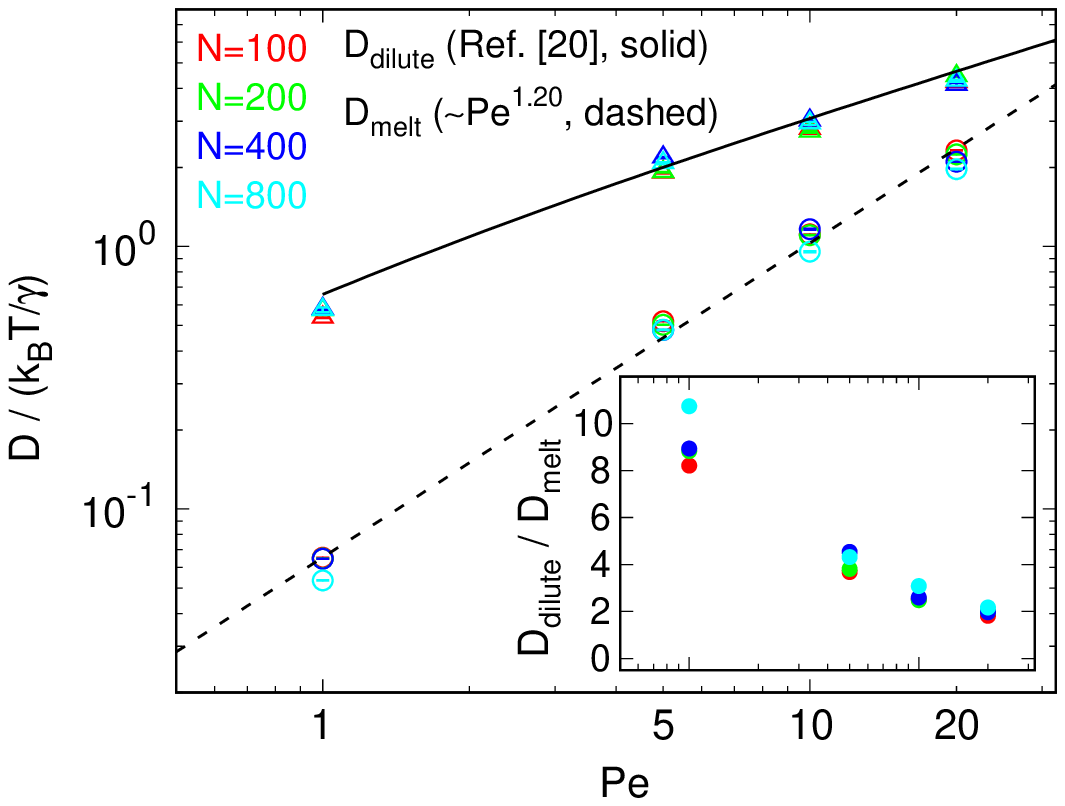}
\caption{
Long-time diffusion coefficient of the centre of mass, $D$ (in units of monomer diffusion coefficient, see Sec.~\ref{sec:SimDetails} in SI), as a function of ${\rm Pe}$.
Symbols are as in Fig.~\ref{fig:R2-BCF-vs-n-scaling-Pe20}.
Lines recapitulating the data correspond to the analytical expression for dilute chains proposed in~\cite{bianco2018globulelike} and a phenomenological power-law fit ($\sim {\rm Pe}^{1.20}$) for melts.
(Inset)
Corresponding ratios, $D_{\rm dilute} / D_{\rm melt}$, between diffusion coefficients in dilute and melt conditions as a function of $\rm Pe$.
}
\label{fig:DiffusionCoeffs}
\end{figure}

Finally, the material properties of the system, such as its viscosity, are of crucial interest for polymer melts and can be connected to the dynamics of the center of mass of the chains~\cite{RubinsteinColbyBook}.
We thus investigate $g_3({\tau})$, the mean-square displacement of the chain centre of mass ${\vec r}_{\rm CM}(t) \equiv \frac1N \sum_{n=1}^N {\vec r}_n(t)$, defined as
\begin{equation}\label{eq:g3}
g_3({\tau}) \equiv \langle \left( {\vec r}_{\rm CM}(t+\tau) - {\vec r}_{\rm CM}(t) \right)^2 \rangle \, ,
\end{equation}
as a function of time $\tau$.
Previous works~\cite{bianco2018globulelike,fazelzadeh2023effects,philipps2022tangentially,li2023nonequilibrium} on dilute polar active polymers has shown that, with respect to passive systems, $g_3(\tau)$ displays a ballistic regime up to the chain relaxation time $\tau_{\rm ee}$, defined as $g_3(\tau_{\rm ee}) = \langle R_{\rm ee}^2 \rangle$, $\langle R_{\rm ee}^2 \rangle$ being the mean-square {\it end-to-end} distance ({\it i.e.}, Eq.~\eqref{eq:R2n} with $n=N-1$); as such, $\tau_{\rm ee}$ is proportional to the end-to-end correlation time, usually employed to characterize $g_3(\tau)$ in polar active polymer models.
At larger times an active diffusive regime ensues $g_3(\tau) = 6 D \tau$, where the active diffusion coefficient $D$ {\it increases} with ${\rm Pe}$ and is {\it independent} of the chain contour length $N$, in contrast with the passive, hydrodynamic-free, predictions~\cite{RubinsteinColbyBook}.
We observe that such features are present also in melt conditions.
Additionally, upon normalizing $g_3({\tau})$ by the corresponding $\langle R_{\rm ee}^2 \rangle$ and $\tau$ by $\tau_{\rm ee}$, data from systems at different values of $N$ and $\mathrm{Pe}$ strikingly collapse onto the same universal curve that is valid for both dilute and melt systems in the ballistic and active diffusive regime (Fig.~\ref{fig:Dynamics-Active} (top panel) and Fig.~\ref{fig:Dynamics-Active-Others} in SI). 
This result is quite remarkable for polymer melts as, according to the classical tube model~\cite{RubinsteinColbyBook}, for $\tau < \tau_{\rm ee}$ single-chain dynamics is expected to be sub-diffusive, $g_3 \sim t^{1/2}$, since motion is confined in a tube-like region, resulting from the {\it topological} barriers ({\it entanglements}) exerted by the surrounding chains.
The absence of this regime suggests that polymer dynamics is entirely controlled by activity, that govers $\tau_{\rm ee}$ and $\langle R_{\rm ee}^2 \rangle$, and entanglements become almost {\it irrelevant} for polymers' dynamics in melts of tangentially active polymers.
In support of this point, we highlight the scaling of $\langle R_{\rm ee}^2 \rangle$ and $\tau_{\rm ee}$ in melt conditions: $\langle R_{\rm ee}^2 \rangle \propto N/\mathrm{Pe}^{0.1}$ and $\tau_{\rm ee} \propto N/\mathrm{Pe}^{1.3}$ (Fig.~\ref{fig:Dynamics-Active}, bottom panels).
Notice, in particular, a difference with the infinite dilution case, where $\langle R_{\rm ee}^2 \rangle \sim N/\mathrm{Pe}^{0.3}$ and $\tau_{\rm ee} \propto N/\mathrm{Pe}$, as well as with passive melts where $\tau_{\rm ee} \sim N^3$~\cite{RubinsteinColbyBook}.
As $D \propto \langle R_{\rm ee}^2\rangle/ \tau_{\rm ee}$, we indeed recover the scaling of $D$ as function of $N$ and $\mathrm{Pe}$, $D \propto N^0 \cdot \mathrm{Pe}^{1.2}$ (see dashed line in Fig.~\ref{fig:DiffusionCoeffs}).
While results for dilute chains are in agreement with the analytical predictions~\cite{bianco2018globulelike} (solid line in Fig.~\ref{fig:DiffusionCoeffs}, see also the brief recapitulation in Sec.~\ref{sec:LongTimeDiffCoeff} in SI), notice that $D$ has a steeper increase in melt; as such 
although dilute chains diffuse systematically faster than in melt, the gap between the two appears to be narrowing for growing $\rm Pe$ (see inset in Fig.~\ref{fig:DiffusionCoeffs}) until the two set-up's become substantially equivalent at sufficiently high values of $\mathrm{Pe}$. 
These findings align with the reported observations that the relaxation times $\tau_{\rm ee}$ for the two set-up's become increasingly close to each other as $\rm Pe$ increases (Fig.~\ref{fig:Dynamics-Active}, bottom right panel).

To summarize, in this work we have investigated the impact of activity on the conformational and dynamical properties of linear chains in melt conditions, as compared to dilute active and passive melt counterparts.
The multi-scale analysis of the conformations highlighted a fundamental length scale, $n_{\rm min}$, which governs both the conformation of the active chains as well the dynamics of the individual monomers. The latter property highlights the ``railway motion'', typical of tangential propulsion, where monomers move along the contour path of the backbone.
Notably, we observe deviations from random walk statistics, particularly for the formation of loops (see Figs.~\ref{fig:R2-vs-n} and~\ref{fig:BCF-vs-n} in SI) within the chains.
Moreover, the asymptotic exponent $\nu$ extracted from $\langle R^2(n)\rangle$ deviates from the random walk value of $\nu = 1/2$. 
While the deviation is small for single chains and both predictions are, overall, compatible with the results reported in~\cite{bianco2018globulelike}, %as we find $\nu \simeq 0.48$, 
chains in the melt deviate significantly reaching a value of $\nu \simeq 0.4$. 
In this respect, some warning needs to be given on attempting to measure $\nu$ from the scaling of the mean-square {\it end-to-end} distance $\langle R_{\rm ee}^2 \rangle$ (namely, Eq.~\eqref{eq:R2n} with $n=N-1$) {\it vs.} $N$.
In fact, as shown in Sec.~\ref{sec:MeanSqE2E} and the corresponding Fig.~\ref{fig:END-END-Active} in SI and in agreement with~\cite{Breoni2023}, the exponent $\nu$ of active chains appears to be $\simeq 0.5$ for {\it both} dilute and melt conditions suggesting that active chains in melt do not remain completely {\it self-similar} (as also pointed out by the mean-square internal distance $\langle R^2(n)\rangle$ in Fig.~\ref{fig:R2-vs-n} in SI~\cite{EquilibriumGlobuleNote}).
At the same time, we have shown that the (apparent) scaling $\langle R_{\rm ee}^2 \rangle \sim N$ plays a role in the characterization of chain dynamics.
We further find that both systems remarkably exhibit the same centre of mass dynamics (see Fig.~\ref{fig:Dynamics-Active}) governed by the length scale $\langle R_{\rm ee}^2\rangle$ and by the time scale $\tau_{\rm ee}$. 
At long times, $\tau \gg \tau_{\rm ee}$, both systems diffuse with a diffusion coefficient, $D$, which increases with ${\rm Pe}$ (see Fig.~\ref{fig:Dynamics-Active}) and is independent on $N$.
However, as mentioned, $D$ has a stronger dependence on $\mathrm{Pe}$ in melts, possibly hinting at the emergence of multi-chain effects that go beyond the railway motion. 
Taken together our results suggest that both chain conformations and dynamics present universal features, that are independent of whether chains are dilute or in melt conditions.
In the latter situation, in particular, the slowing effects due to the unavoidable topological constraints (entanglements) between distinct chains  -- a fundamental property of their passive counterparts -- become negligible when chains are subjected to tangential active forces.
These results align with experimental observations~\cite{deblais2020rheology} on active worms, which showed a decrease in system viscosity with activity, and a milder dependence on concentration compared to passive entangled systems.
Moreover, as recently shown~\cite{Breoni2023}, for tangentially propelled entangled chains the plateau modulus, $G_0$, that encodes the elastic response of the system~\cite{RubinsteinColbyBook}, depends on the active forces in the systems and is not anymore simply proportional~\cite{RubinsteinColbyBook} to the mean concentration of entanglement lengths as in equilibrium conditions. %{\it i.e.}$G_0 \simeq \rho \kappa_BT / N_e$.

%%%
\section*{Supporting Information}
%%%
Complementary details on: polymer model, molecular dynamics computer simulations and preparation and equilibration of polymer configurations, single-chain diffusion and mean-square end-to-end distance.
Supporting table for scaling parameters $n_{\rm min}$ and $\langle R^2(n_{\rm min}) \rangle$.
Additional figures on: polymer structure (internal distances and bond-vector correlations), polymer dynamics (mean-square displacements for monomers and chain centre of mass).

\begin{acknowledgments} 
This work was supported by a STSM Grant from COST Action CA17139 (eutopia.unitn.eu) funded by COST (www.cost.eu). E. Locatelli acknowledges support from the MIUR grant Rita Levi Montalcini. 
\end{acknowledgments}
%

%%%
\bibliography{bibliografia}

\begin{thebibliography}{44}
\bibitem [{Equ()}]{EquilibriumGlobuleNote}%
  \BibitemOpen
  \href@noop {} {}\bibinfo {note} {This should not be too surprising. An
  analogous situation occurs for flexible polymers that are densely packed in a
  spherical cavity. For increasing chain length $N$ the radius of the sphere
  must also increase to keep the density fixed and, trivially, $\langle R_{\rm
  ee}^2 \rangle \sim N^{2/3}$. On the other hand, on length scales smaller than
  the radius of the sphere, the local structure of the system is no different
  than the one of an ordinary polymer melt, implying $\nu=1/2$.}\BibitemShut
  {Stop}%
\bibitem [{Sys()}]{SystemsIniPrep}%
  \BibitemOpen
  \href@noop {} {}\bibinfo {note} {Note that, only for the preparation of the
  initial states, we have adopted the conventional~\cite{KremerGrest-JCP1990}
  Kremer-Grest values for the parameters in Eq.~\eqref{eq:Ufene}, namely:
  $\epsilon = 1\, \kappa_BT$ and $k=30 \, \kappa_B T/\sigma^2$.}\BibitemShut
  {Stop}%
\end{thebibliography}
%%%

%%%
\clearpage

%\appendix

\title{}
%%%%
\widetext
\clearpage
\begin{center}
\textbf{\Large  \\ \vspace*{1.5mm} -- Supporting Information -- \\ Universal time and length scales of polar active polymer melts} \\
\vspace*{5mm}
Mattia Alberto Ubertini, Emanuele Locatelli, Angelo Rosa
%Authors
\vspace*{10mm}
\end{center}
\balancecolsandclearpage
%%%

%%%
\onecolumngrid
%%%

%
\tableofcontents
\clearpage
%

%%%%%%%%% Prefix a "S" to all equations, figures, tables and reset the counter %%%%%%%%%%
\setcounter{equation}{0}
\setcounter{figure}{0}
\setcounter{table}{0}
\setcounter{page}{1}
\setcounter{section}{0}
\setcounter{page}{1}
\makeatletter
\renewcommand{\theequation}{S\arabic{equation}}
\renewcommand{\thefigure}{S\arabic{figure}}
\renewcommand{\thetable}{S\arabic{table}}
\renewcommand{\thesection}{S\arabic{section}}
\renewcommand{\thepage}{S\arabic{page}}
%%%

%\tableofcontents
%\addcontentsline{toc}{chapter}{Appendices}
%\addtocontents{toc}{\protect\setcounter{tocdepth}{1}}
%\input{myTOC.toc}

In this Supporting Information file, we provide more specific details on the polymer model used (Sec.~\ref{sec:PolymModel-Active}), on the compositions of the considered polymer systems and technical details on the molecular dynamics computer simulations (Sec.~\ref{sec:SimDetails}) and on the initial preparation and following equilibration of the systems (Sec.~\ref{sec:IniConfig}).
We conclude with a brief overview on the scaling properties of the mean-square end-to-end distance for our polymers (Sec.~\ref{sec:MeanSqE2E}) and a short recap (Sec.~\ref{sec:LongTimeDiffCoeff}) of the theory for active polymer model diffusion first discussed in Refs.~\cite{bianco2018globulelike,fazelzadeh2023effects} and used in the main text.

%%%
\section{Polymer model and methods}\label{sec:PolymerModelMethods}
%%%

%%%
\subsection{Numerical model for active polymers}\label{sec:PolymModel-Active}
%%%
Chain connectivity and monomer-monomer interactions for dilute and melt linear chains are accounted for by a suitably modified version of the classical polymer model by Kremer and Grest~\cite{KremerGrest-JCP1990}.
Specifically, excluded volume interactions between beads (including consecutive ones along the contour length of the chains) are described in terms of the shifted and truncated Lennard-Jones (LJ) potential:
\begin{equation}\label{eq:LJ}
U_{\rm LJ}(r) = \left\{
\begin{array}{lr}
4 \epsilon \left[ \left(\frac{\sigma}{r}\right)^{12} - \left(\frac{\sigma}{r}\right)^6 + \frac14 \right] & \, r \le r_c \\
0 & \, r > r_c
\end{array} \right. \, ,
\end{equation}
where $r$ denotes the spatial separation between the bead centers.
By denoting with $\kappa_B$ the Boltzmann constant, the energy scale is set to $\epsilon = 20 \, \kappa_BT$ (against the standard $\epsilon = 1\hspace{0.1cm}\kappa_BT$~\cite{KremerGrest-JCP1990}) where $T$ and $\sigma$ are, respectively, the units of temperature and length in our simulation.
The unit of energy is thus taken as the thermal energy, {\it i.e.} $\kappa_B T=1$.
Further, nearest-neighbour monomers along the contour of the chains are connected by the finitely extensible nonlinear elastic (FENE) potential, given by:
\begin{equation}\label{eq:Ufene}
U_{\rm FENE}(r) = \left\{
\begin{array}{lcl}
-0.5kR_0^2 \ln\left(1-(r / R_0)^2\right) & \ r\le R_0 \\ \infty & \
r> R_0 &
\end{array} \right. \, ,
\end{equation}
where $k = 30 \epsilon / \sigma^2 = 600 \, \kappa_B T / \sigma^2$ is the spring constant and $R_{0}=1.5\sigma$ is the maximum extension of the elastic FENE bond~\cite{WhyFENEnote}.
Finally, polymer activity is taken into account by imposing that the $i$-th monomer of each chain of spatial coordinates $\vec r_i$ (for $i=2, ..., N-1$, with $N$ being the total number of monomers of the chain) is subject to the active force $\vec F_i$~\cite{bianco2018globulelike}:
\begin{equation}\label{eq:ActiveForce}
\vec F_i = f_a \frac{\vec r_{i+1} - \vec r_{i-1}}{|\vec r_{i+1} - \vec r_{i-1}|} \, ,
\end{equation}
of constant magnitude $f_a$ and instantaneous orientation directed along the tangent to the polymer chain at $\vec r_i$.
Notice that the first and the last monomers of each chain are (conventionally, see~\cite{bianco2018globulelike}) excluded by the active perturbation because those monomers have only one, instead of two, neighbor along the chain.
As described in the main paper, the relative importance of the active forces {\it vs.} the thermal ones is quantified in term of the so called P\'eclet number $\rm Pe$, defined as
\begin{equation}\label{eq:DefinePecletNumber}
{\rm Pe} \equiv \frac{f_a \sigma}{\kappa_B T} \, ,
\end{equation}
We consider values ${\rm Pe}=1, 5, 10, 20$ and, for comparison, the purely passive case ${\rm Pe}=0$.

%%%
\subsection{Simulation details}\label{sec:SimDetails}
%%%
We consider monodisperse melts of $M$ linear polymer chains, each chain being made of $N$ monomers.
Specifically, we consider systems with compositions
$(M, N) = (1000, 100)$, $(500, 200)$ and $(250, 400)$ ({\it i.e.}, $M\times N = 100,\!000$ monomers in total)
and $(M, N) = (250, 800)$ ({\it i.e.}, $M\times N = 200,\!000$ monomers in total). 
As in the original work by Kremer and Grest~\cite{KremerGrest-JCP1990}, we maintain a fixed monomer density of $\rho = 0.85 \sigma^{-3}$ for all polymer compositions.

As mentioned in the main paper, we compare simulation results for active polymer melts to those for single self-avoiding active polymers in dilute conditions.
In the latter case we simulate the same systems as the ones of the set-up's described above, with deactivated {\it inter}-chain LJ (see Eq.~\eqref{eq:LJ}) excluded volume interactions.
This allows us to effectively simulate replicas of linear chains in dilute conditions.

The static and kinetic properties of chains are studied using fixed-volume and constant-temperature {\it molecular dynamics} (MD) simulations with implicit solvent and periodic boundary conditions.
MD simulations are performed by using the LAMMPS package~\cite{thompson2022lammps}.
By introducing the MD time-unit $\tau_{\rm MD} = \sigma \sqrt{m / \kappa_B T}$, we integrate the equations of motion by using the velocity-Verlet algorithm. 
We set $\Delta t$, the integration time step of the algorithm, as the following:
(i) $\Delta t = 0.01 \, \tau_{\rm MD}$ for passive polymers ({\it i.e.}, $\rm{Pe}=0$)
and
(ii) $\Delta t = 0.001 \, \tau_{\rm MD}$ for active systems with ${\rm Pe} >0$ (in order to prevent accidental strand-crossings, especially in the regime of high active forces).
Finally, in order to ensure the overdamped regime~\cite{fazelzadeh2023effects}, the friction coefficient $\gamma$ is $=20 \, \tau^{-1}_{MD}$.

%%%
\subsection{Preparation of initial chain configurations and check for equilibration}\label{sec:IniConfig}
%%%
Preparation of melts of linear chains poses no particular technical problem.
Chains up to $N=400$ were initially arranged inside the simulation box and the system is passively ({\it i.e.}, active forces at this stage are turned off) let towards complete equilibration (defined as after the chains have drifted from their original positions for several times their own root-mean-square end-to-end distance, $\sqrt{ \langle R_{\rm ee}^2 \rangle }$ (where $\langle R_{\rm ee}^2 \rangle$ is Eq.~\eqref{eq:R2n} in the main text with $n=N-1$).
For the case with $N=800$ this procedure becomes tediously long, so we have prepared first the system on the FCC lattice and, then, let it partially equilibrate by means of the efficient kinetic Monte Carlo algorithm used in~\cite{ubertini2021computer,ubertini2022entanglement}.
After completing this step, we move the whole system back off-lattice again and let it evolve by standard MD: being the system highly entangled for the polymer contour length considered, we performed one last equilibration check by verifying that the main structural properties of the chains are effectively independent of the initial preparation~\cite{SystemsIniPrep}.

Following this initial set-up, the polymer conformations become the starting point for the production runs of systems in the presence of active force ({\it i.e.}, for ${\rm Pe} > 0$) and, also, for their passive counterparts (${\rm Pe}=0$).
The typical production run consists of $10^6 \, \tau_{\rm MD}$-units.

%%%
\section{More on single-chain properties}\label{sec:MoreResults}
%%%

%%%
\subsection{Long-time diffusion coefficient}\label{sec:LongTimeDiffCoeff}
\begin{figure*}
%\begin{center}
$$
\begin{array}{cc}
{\rm Pe}=1 & {\rm Pe}=5 \\
\includegraphics[width=0.50\textwidth]{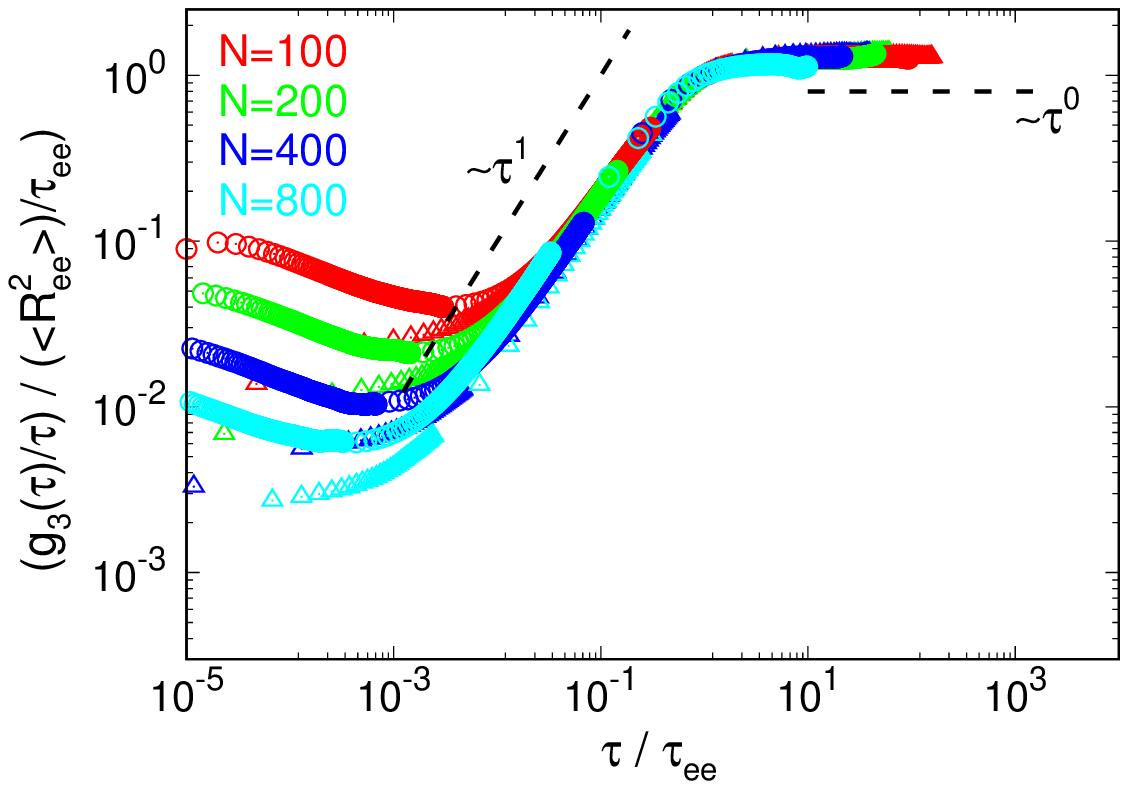} & \includegraphics[width=0.50\textwidth]{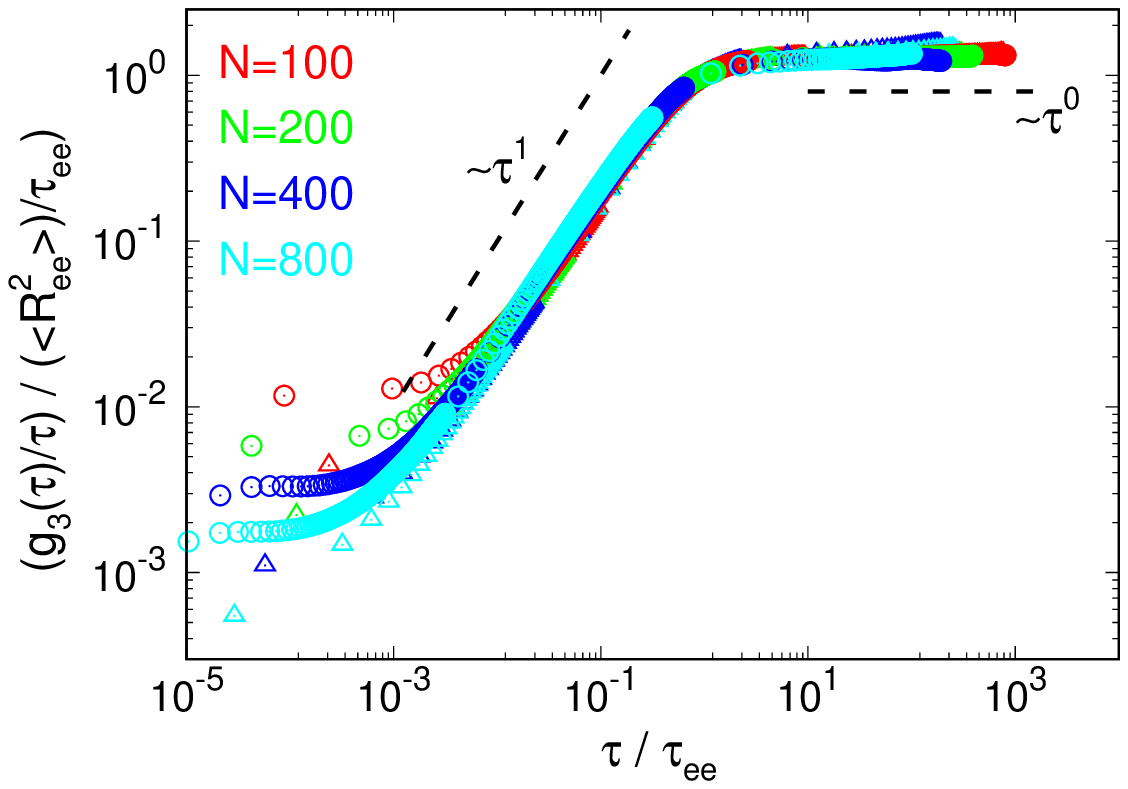} \\
\\
\multicolumn{2}{c}{ {\rm Pe}=10 } \\
\multicolumn{2}{c}{ \includegraphics[width=0.50\textwidth]{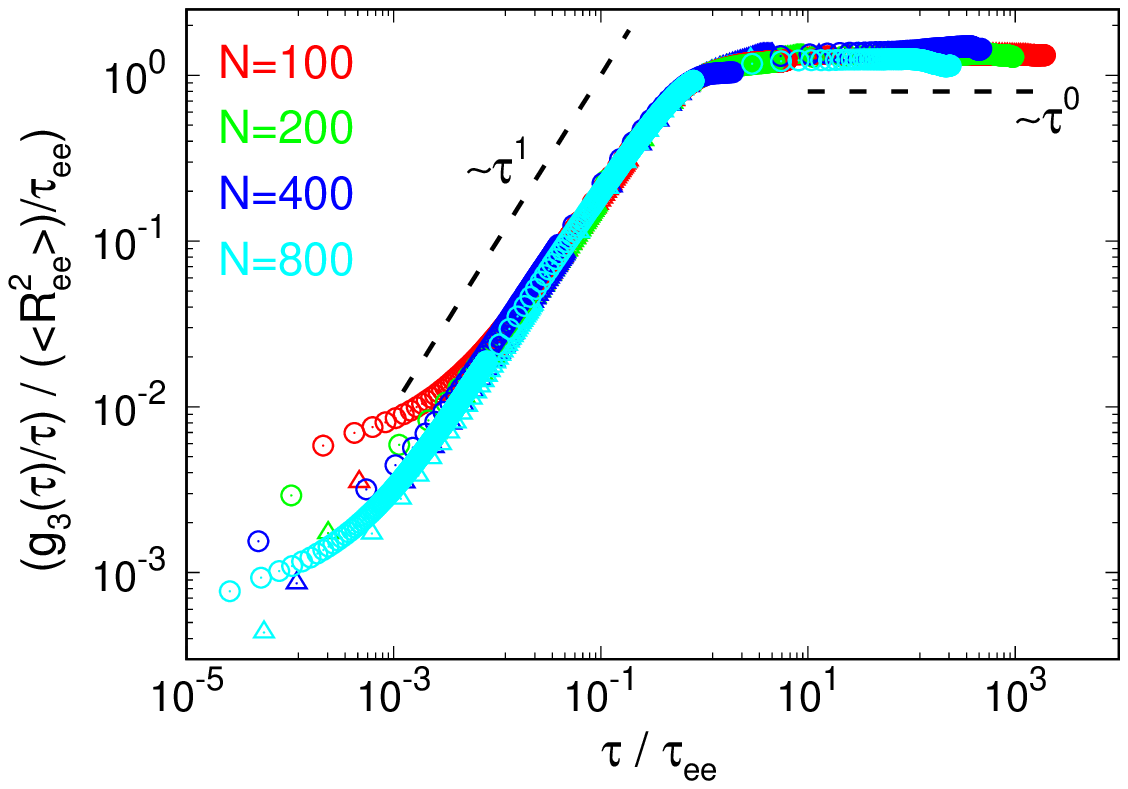} }
\end{array}
$$
%\end{center}
\caption{
Mean-square displacement of the chain centre of mass per unit time, $g_3(\tau) / \tau$, normalized by $\langle R_{\rm ee}^2 \rangle / \tau_{\rm ee}$ as a function of normalized time $\tau / \tau_{\rm ee}$. 
Results for polymers in dilute ($\bigtriangleup$) and melt ($\circ$) conditions, for different contour lengths $N$ (see legend) and for P\'eclet numbers ${\rm Pe}=1, \, 5, \, 10$ (for ${\rm Pe}=20$, see top panel in Fig.~\ref{fig:Dynamics-Active} in the main paper).
}
\label{fig:Dynamics-Active-Others}
\end{figure*}

%%%
The long time diffusion coefficient of isolated tangentially active polymers can be computed analytically~\cite{bianco2018globulelike,fazelzadeh2023effects}.
We start from the diffusion coefficient for an Active Brownian Particle,
\begin{equation}\label{eq:Dabp}
D= D_t + \frac{\tau_r \, v_a^2}{2 d} = D_t + \frac{\tau_r \left( F_a/\gamma \right)^2}{2 d} \, ,
\end{equation}
where $v_a = f_a/\gamma$ is the self-propulsion velocity and $D_t = \kappa_B T/\gamma$; $d$ is the dimensionality of the system (here, $d=3$).
For a tangentially active polymer:
(i)
$\gamma = N\gamma_0$, $\gamma_0$ being the friction coefficient of a single monomer;
(ii)
$F_a = f_a R_{\rm ee}/\sigma$, with $f_a = \frac{\kappa_B T}{\sigma}{\rm Pe}$ as in Eq.~\eqref{eq:DefinePecletNumber}, $R_{\rm ee}$ being the (root mean-square) end-to-end distance and $\sigma$ the mean bond length (Sec.~\ref{sec:PolymModel-Active});
(iii)
$\tau_r = \tau_{\rm ee}$ corresponds to the chain relaxation time (defined in the main text). 
If $f_a$ is sufficiently large, we can disregard the passive contribution and write
\begin{equation}\label{eq:Dtheo}
D = \frac{\tau_{\rm ee}}{2d} \, \left( \frac{\mathrm{Pe}}{N} \frac{\kappa_B T}{\gamma_0} \frac{R_{\rm ee}}{\sigma} \right)^2 = \frac{1}{2d} \frac{\tau_{\rm ee} D_0}{\sigma^2} \, \frac{\mathrm{Pe}^2}{N^2}  \frac{R^2_{\rm ee}}{\sigma^2} D_0 \, .
\end{equation} 
We now consider what is reported in the top panel of Fig.~\ref{fig:Dynamics-Active} in the main text and Fig.~\ref{fig:Dynamics-Active-Others} here, {\it i.e.} $(g_3(\tau)/\tau)/(R_{\rm ee}^2/\tau_{\rm ee})$.
We focus on the long time diffusion: we have by definition that $g_3(\tau) = 2d D \tau$, from which $g_3(\tau)/\tau = 2d D$.
As such, the data reported in Fig.~\ref{fig:Dynamics-Active} in the main text and Fig.~\ref{fig:Dynamics-Active-Others} here show that 
\begin{equation}\label{eq:eqfig4}
(g_3(\tau)/\tau)/(R_{\rm ee}^2/\tau_{\rm ee}) = (2d D)/(R_{\rm ee}^2/\tau_{\rm ee}) \simeq 1.5 \, .
\end{equation}
From Eq.~\eqref{eq:Dtheo} we get 
\begin{equation}
\frac{2d D}{R_{\rm ee}^2/\tau_{\rm ee}} = \left( \frac{\tau_{\rm ee} D_0}{\sigma^2} \right)^2 \, \frac{\mathrm{Pe}^2}{N^2} = \left( \tau_0 \frac{N}{\mathrm{Pe}} \right)^2 \, \frac{\mathrm{Pe}^2}{N^2} = \tau_0^2 \, ,
\end{equation}
where $D_0$ is the passive diffusion coefficient of a single monomer and we have used the relation $\tau_{\rm ee} D_0/\sigma^2 = \tau_0 N / {\rm Pe}$, shown to be valid in dilute conditions for the characteristic relaxation time $\tau_{\rm ee}$ in Fig.~\ref{fig:Dynamics-Active} of the main text.
As $D_0=\kappa_B T/\gamma = 0.05$ in this work, from Fig.~\ref{fig:Dynamics-Active} we get $\tau_0 = 1.05$.
As such, the analytical prediction is $(2d D)/(R_{\rm ee}^2 / \tau_e) = 1.1$.

%%%
\subsection{Mean-square end-to-end distance}\label{sec:MeanSqE2E}
\begin{figure*}%[!h]
\includegraphics[width=0.95\textwidth]{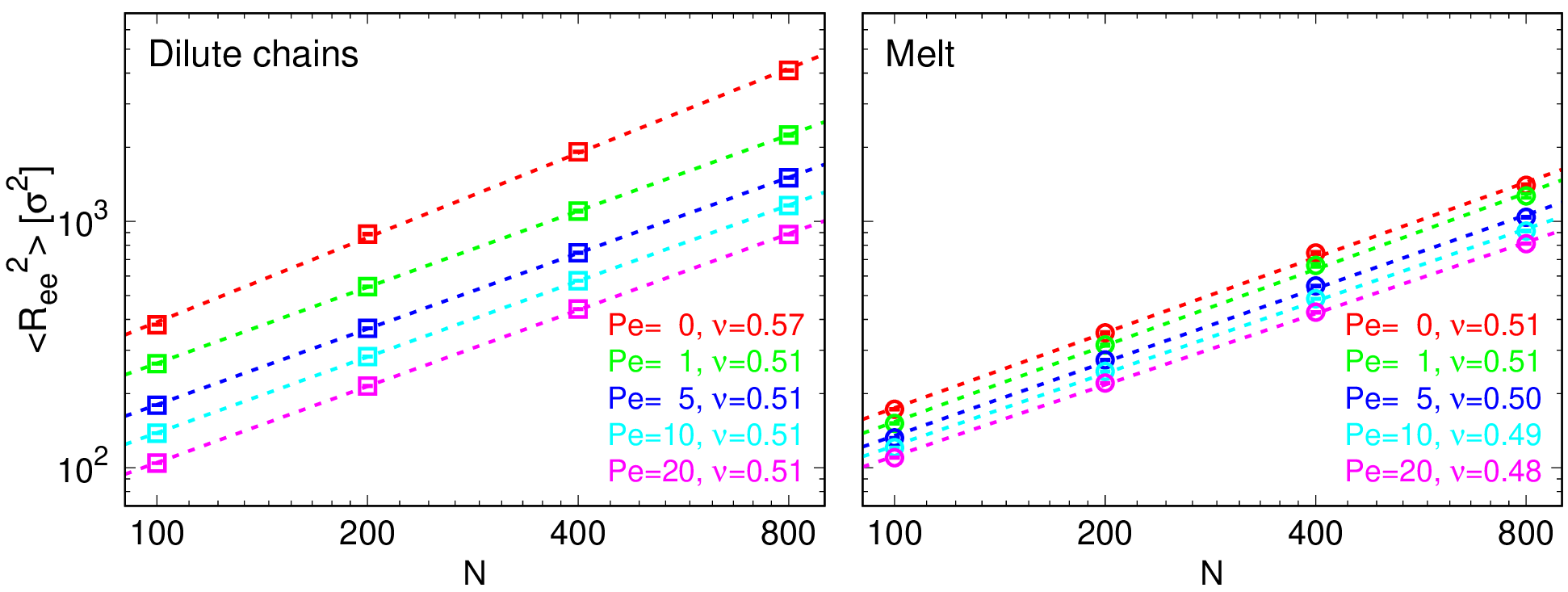}
\caption{
Polymer mean-square end-to-end distance, $\langle R_{\rm ee}^2 \rangle$ (Eq.~\eqref{eq:R2n} in the main paper, with $n=N-1$), as a function of the total chain contour length $N$ and P\'eclet number $\rm Pe$ (Eq.~\eqref{eq:DefinePecletNumber}).
Dashed lines are best fits to simple power-law behavior $=b^2 N^{2\nu}$ (see legends, for the results).
}
\label{fig:END-END-Active}
\end{figure*}

%%%
Results for $\langle R_{\rm ee}^2 \rangle$ are reported in Fig.~\ref{fig:END-END-Active}. 
The passive case (${\rm Pe} =0$) behaves as expected, dilute linear chains exhibit swelling compared to chains within the melt, with their sizes scaling differently.
Specifically, the scaling exponent for dilute chains is $\nu = 0.588$~\cite{RubinsteinColbyBook}, while in the melt, linear chains follow random walk statistics~\cite{RubinsteinColbyBook} and their sizes scale with $\nu = 0.5$. 
Notably, in both systems, as $\rm{Pe}$ grows, the polymers crumple, a feature already seen for dilute active linear chains~\cite{bianco2018globulelike}.
Remarkably, at high P\'eclet number $\simeq 10/20$, across all considered polymer lengths $N$, $\langle R_{\rm ee}^2 \rangle$ for both dilute and melt cases converge to similar values. 
Moreover, always at high activity, the scaling of $\langle R_{\rm ee}^2 \rangle$ appears to align with ideal behavior, exhibiting a scaling exponent of $\nu = 0.5$ for both cases.

%%%
\clearpage
\section*{Supporting Table \& Figures}
\clearpage
%%%

%
\begin{table*}%[h]
\resizebox{\columnwidth}{!}{
\begin{tabular}{ccccccccccccccccccccccc}
\hline
\hline
\\
& & & \multicolumn{9}{c}{ Dilute } & & & \multicolumn{9}{c}{ Melt } \\
\cline{4-12} \cline{15-23}
\, ${\rm Pe}$ \, & & & \, $n_0$ \, & & \, $\alpha$ \, & & \, $R_0^2$ \, & & \, $\beta$ \, & & \, $\beta/\alpha$ \, & & & \, $n_0$ \, & & \, $\alpha$ \, & & \, $R_0^2$ \, & & \, $\beta$ \, & & \, $\beta/\alpha$ \, \\
\hline
$1$ & & & $8.2 \pm 1.7$ & & $0.36 \pm 0.02$ & & $17.4 \pm 4.1$ & & $0.45 \pm 0.02$ & & $1.2 \pm 0.1$ & & & $5.2 \pm 1.6$ & & $0.37 \pm 0.03$ & & $4.7 \pm 1.7$ & & $0.63 \pm 0.02$ & & $1.7 \pm 0.2$\\
$5$ & & & $3.3 \pm 0.2$ & & $0.330 \pm 0.007$ & & $4.9 \pm 0.4$ & & $0.435 \pm 0.006$ & & $1.32 \pm 0.05$ & & & $3.3 \pm 0.9$ & & $0.37 \pm 0.03$ & & $2.8 \pm 1.3$ & & $0.63 \pm 0.04$ & & $1.7 \pm 0.2$\\
$10$ & & & $2.4 \pm 0.2$ & & $0.311 \pm 0.008$ & & $3.0 \pm 0.3$ & & $0.433 \pm 0.008$ & & $1.39 \pm 0.06$ & & & $3.1 \pm 0.8$ & & $0.35 \pm 0.02$ & & $2.7 \pm 1.1$ & & $0.59 \pm 0.04$ & & $1.7 \pm 0.2$\\
$20$ & & & $2.4 \pm 0.1$ & & $0.259 \pm 0.005$ & & $2.7 \pm 0.1$ & & $0.396 \pm 0.006$ & & $1.53 \pm 0.05$ & & & $2.9 \pm 0.5$ & & $0.32 \pm 0.02$ & & $2.6 \pm 0.7$ & & $0.55 \pm 0.03$ & & $1.7 \pm 0.2$\\
\hline
\hline
\end{tabular}
}
\caption{
Fit parameters for the empirical power-laws $n_{\rm min} = n_0 \! \left(\frac{N}{n_0}\right)^{\!\alpha}$ and $\langle R^2(n_{\rm min}) \rangle = R_0^2 \left(\frac{N}{n_0}\right)^{\!\beta}$ (see main text for details) for dilute and melt systems and at P\'eclet number $\rm Pe$.
By combining these two power-laws, one get the scaling $\langle R^2(n_{\rm min}) \rangle \sim n_{\rm min}^{\beta/\alpha}$: the exponent $\beta / \alpha$ has been calculated from the corresponding values for $\alpha$ and $\beta$.
}
\label{tab:nMinFitParams}
\end{table*}
\begin{figure*}%[!h]
\centering
\includegraphics[width=0.62\textwidth]{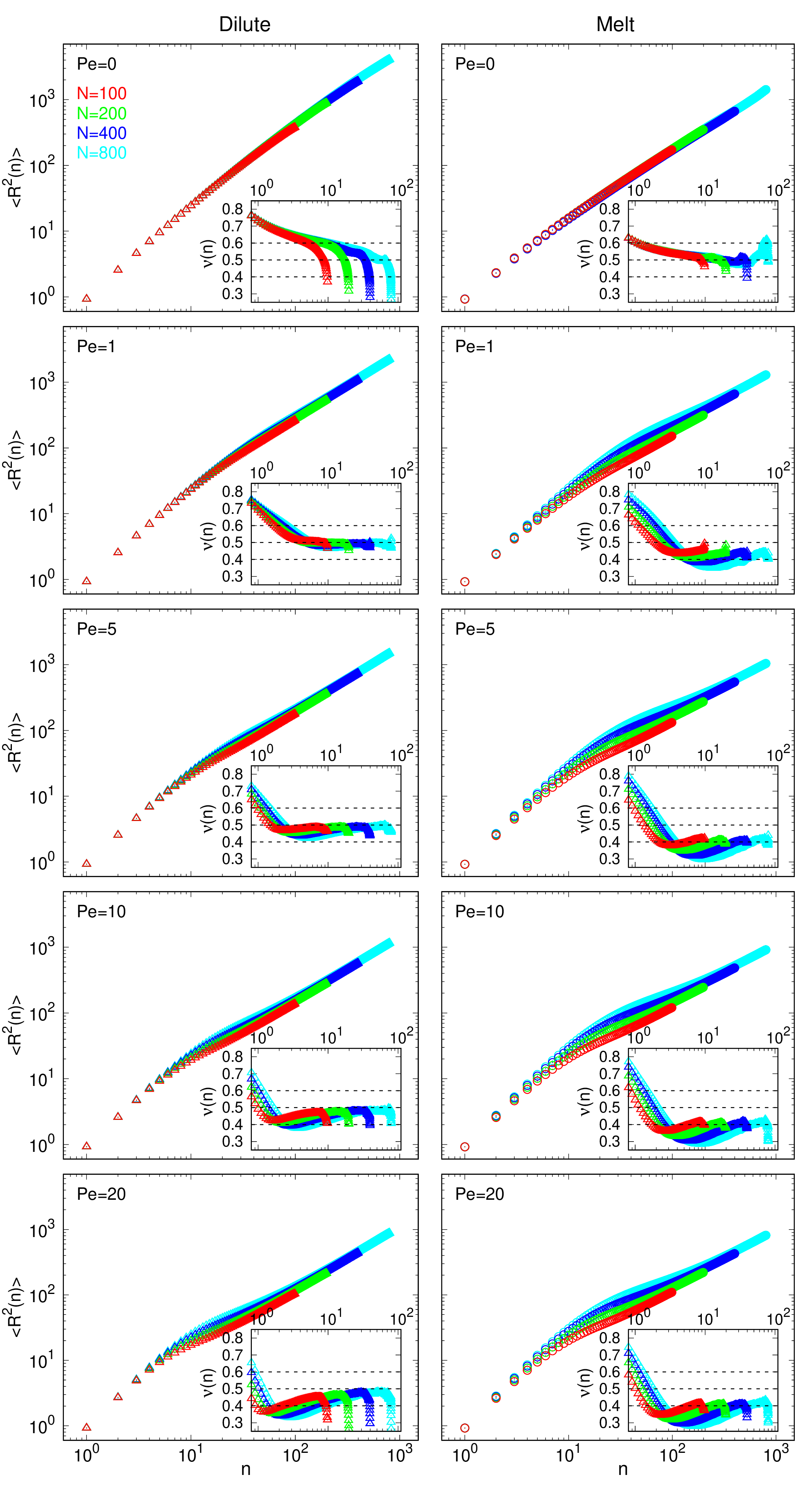}
\caption{
Mean-square internal distance, $\langle R^2(n) \rangle$ (Eq.~\eqref{eq:R2n} in the main paper), as a function of contour length separation, $n$, and (insets) corresponding local scaling exponents $\nu(n)$ (Eq.~\eqref{eq:DefineNu} in the main paper).
Left- and right-hand-side panels are for dilute and melt systems.
The dashed lines in the insets are for reference values $\nu=0.4$, $0.5$ and $0.6$.
Increasing P\'eclet numbers are from top to bottom, while different colors are for different total chain lengths $N$ (see legend).
}
\label{fig:R2-vs-n}
\end{figure*}
\begin{figure*}
\includegraphics[width=0.65\textwidth]{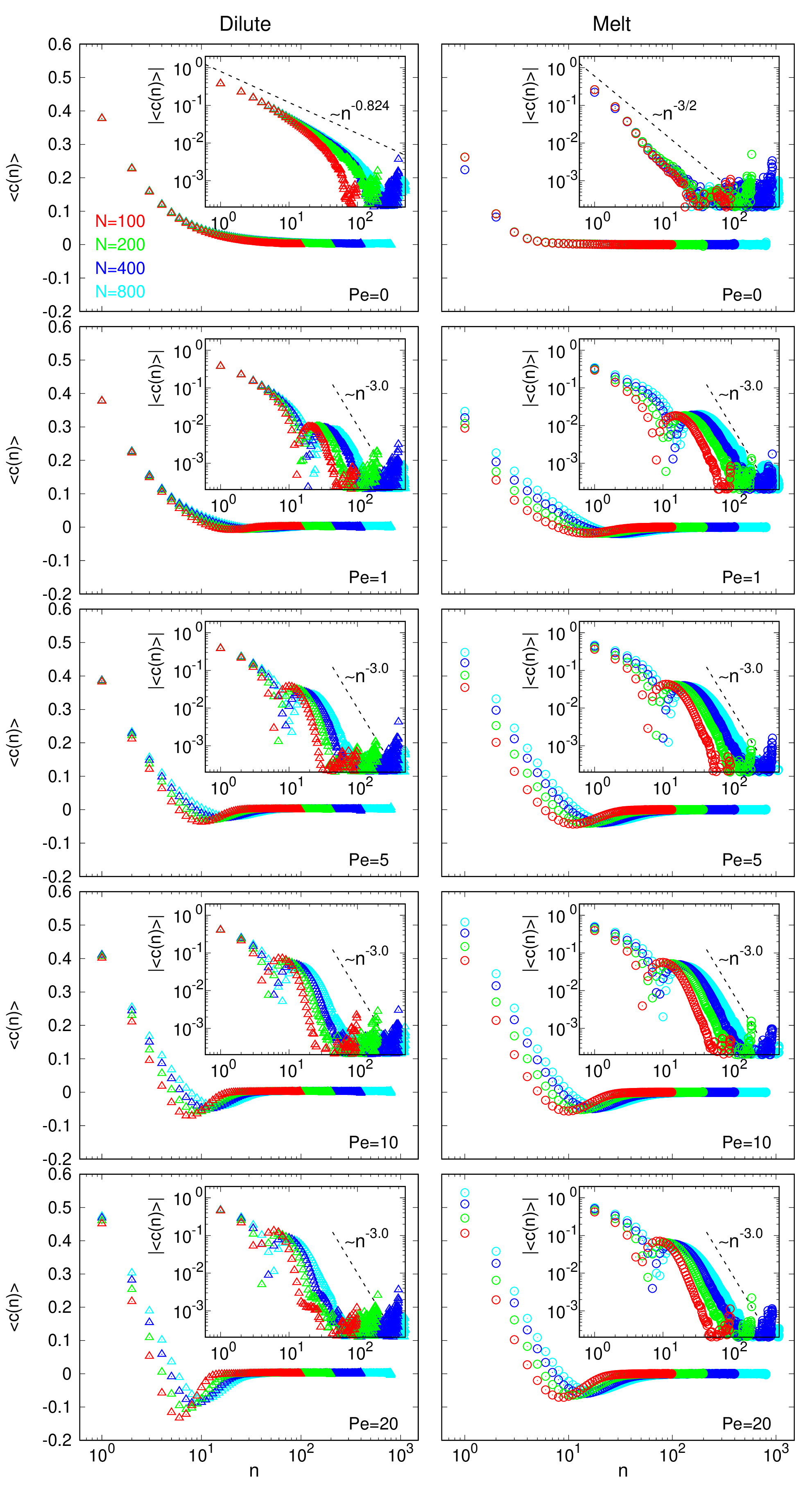}
\caption{
Bond-vector correlation function, $\langle c(n)\rangle$ (Eq.~\eqref{eq:tgtgCorrel} in the main paper), as a function of contour length separation, $n$, in log-lin representation (main panels) and corresponding norm, $|\langle c(n)\rangle|$, in log-log representation (insets).
Panels ordering, symbols and color code are as in Fig.~\ref{fig:R2-vs-n}.
Power-law decays in the ``${\rm Pe}=0$''-insets correspond to the known~\cite{WittmerPRL2004} behaviors in dilute and melt polymers. 
The ``$\sim n^{-3}$''-decay for active (${\rm Pe} >0$) systems is for guiding the eye and purely indicative.
}
\label{fig:BCF-vs-n}
\end{figure*}
\begin{figure*}
$$
\begin{array}{c}
{\rm Pe}=1 \\
\includegraphics[width=0.50\textwidth]{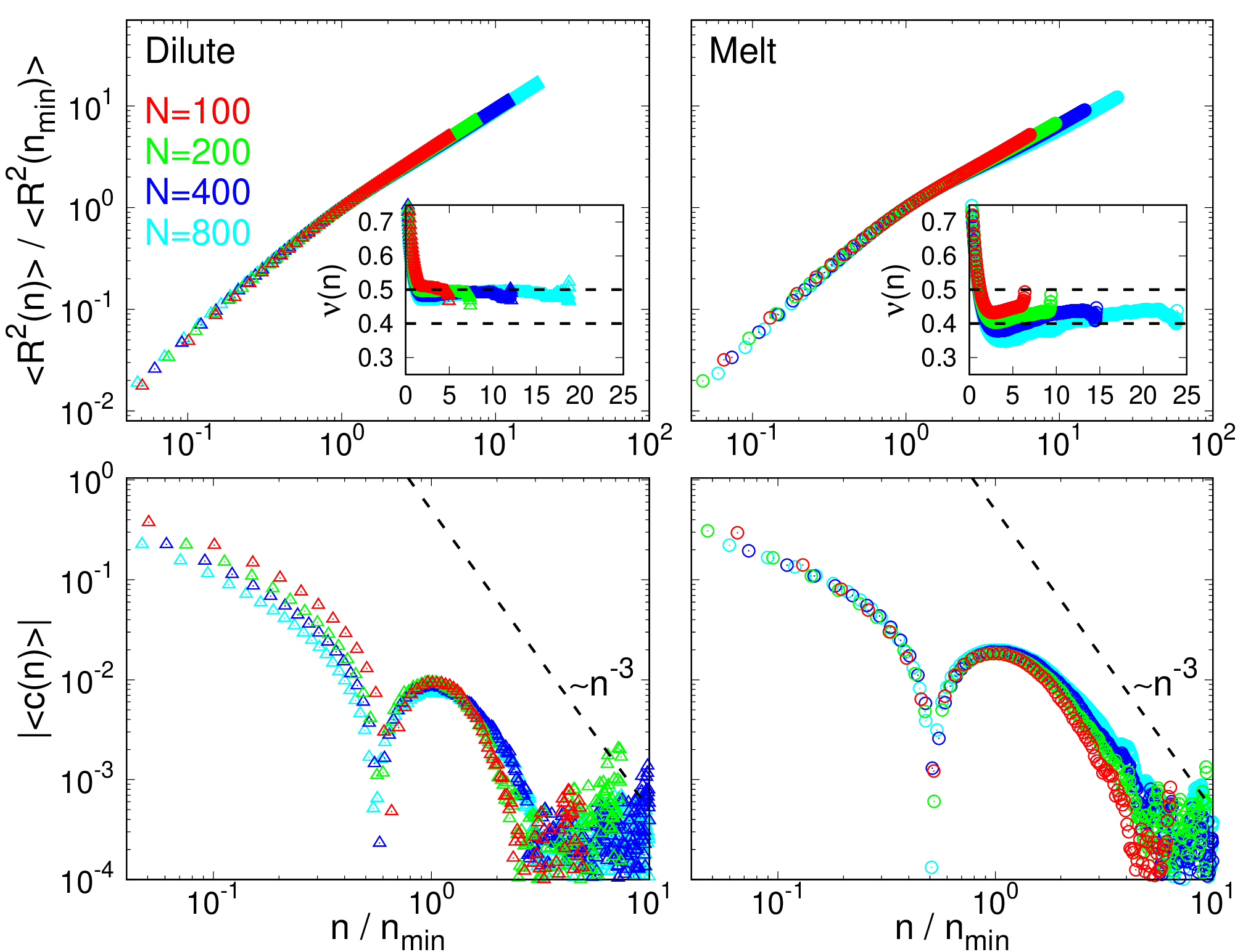} \\
\\
\\
{\rm Pe}=5 \\
\includegraphics[width=0.50\textwidth]{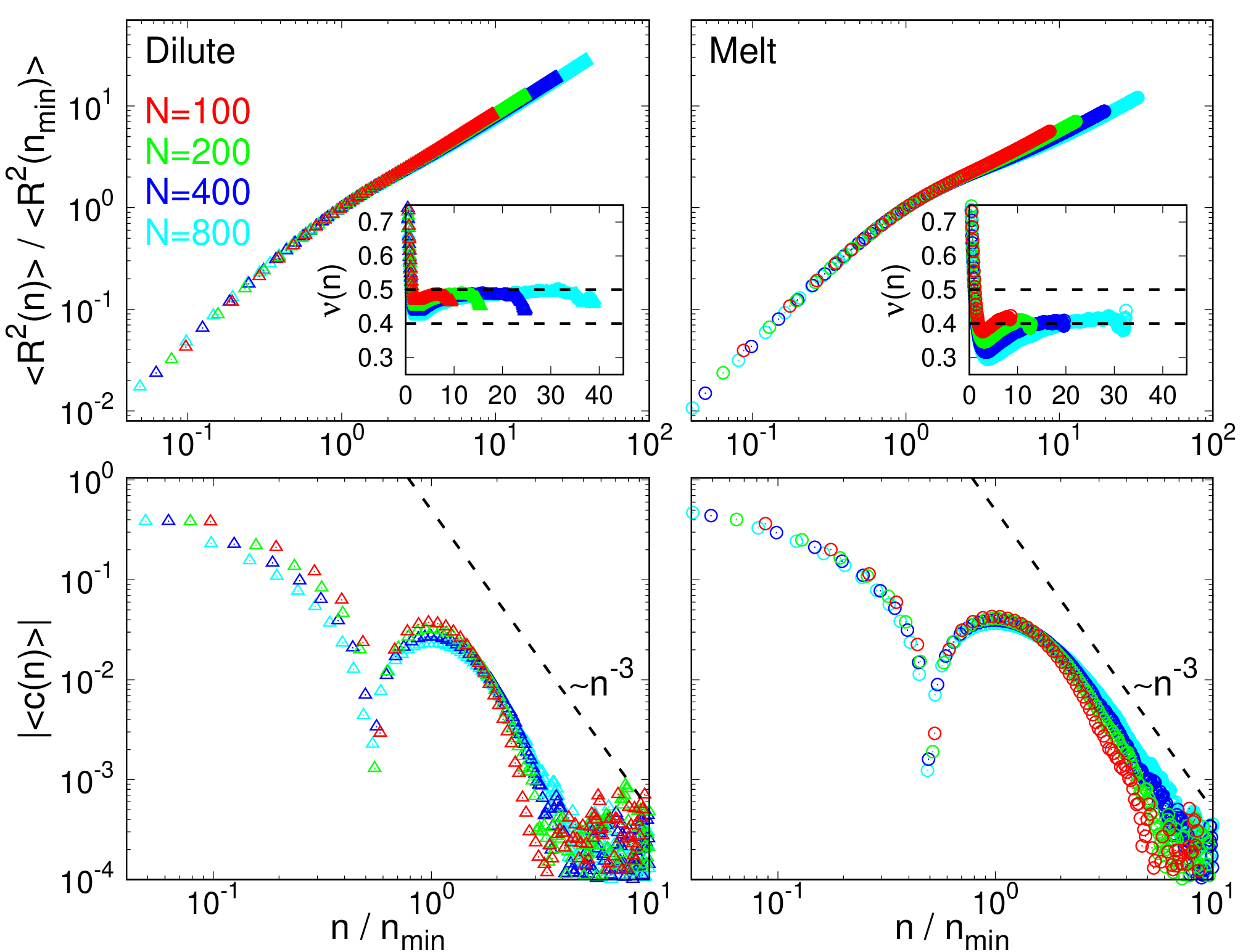} \\
\\
\\
{\rm Pe}=10 \\
\includegraphics[width=0.50\textwidth]{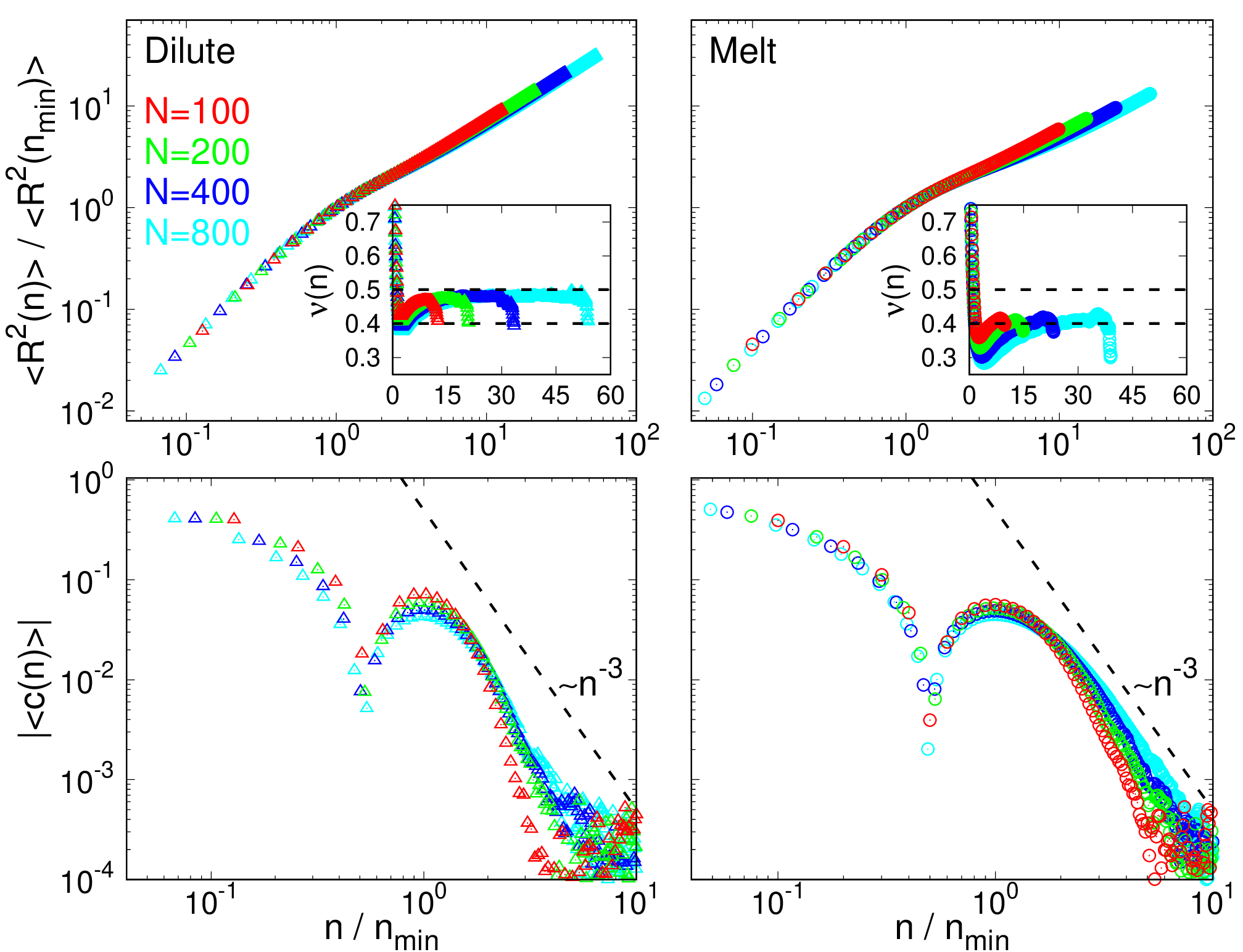}
\end{array}
$$
\caption{
Spatial structure of active polymer chains of contour length $N$ (different colors are for different $N$'s, see legend) and for P\'eclet numbers ${\rm Pe}=1, \, 5, \, 10$.
Notation, symbols and color code are as in Fig.~\ref{fig:R2-BCF-vs-n-scaling-Pe20} in the main paper.
}
\label{fig:R2-BCF-vs-n-scaling-Pe>0}
\end{figure*}
\begin{figure*}
$$
\begin{array}{c}
{\rm Pe}=1 \\
\includegraphics[width=0.75\textwidth]{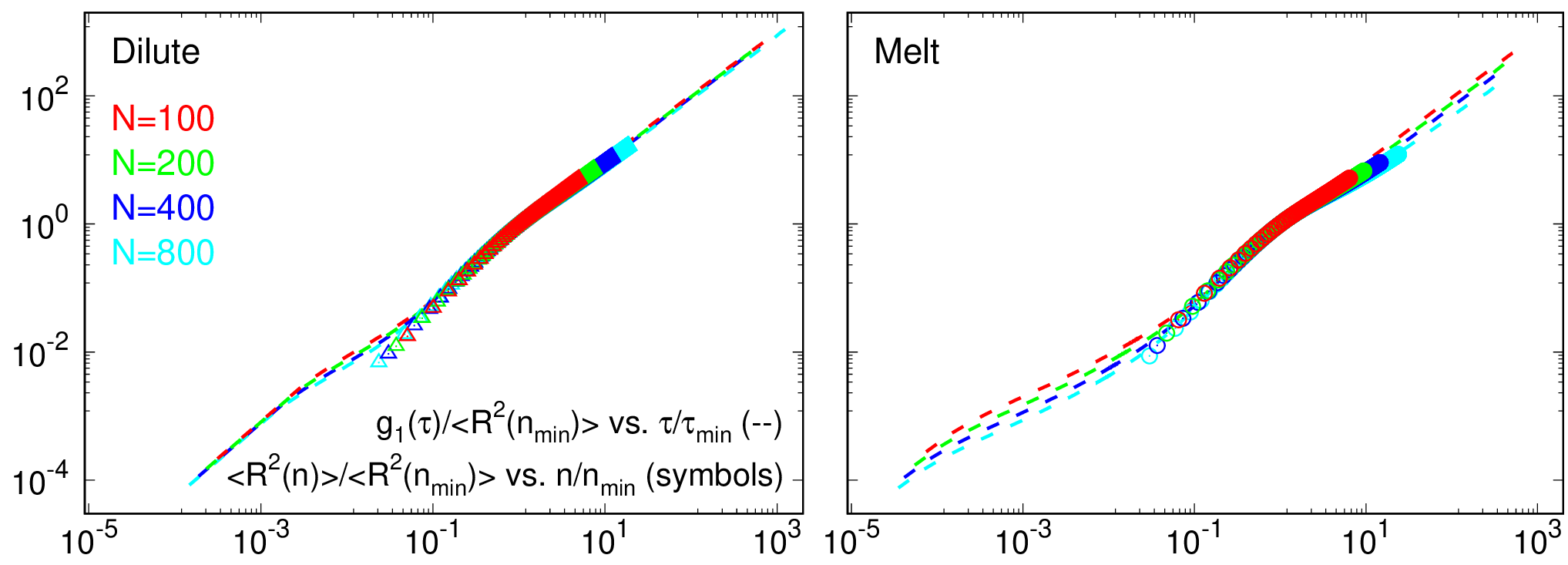} \\
\\
\\
{\rm Pe}=5 \\
\includegraphics[width=0.75\textwidth]{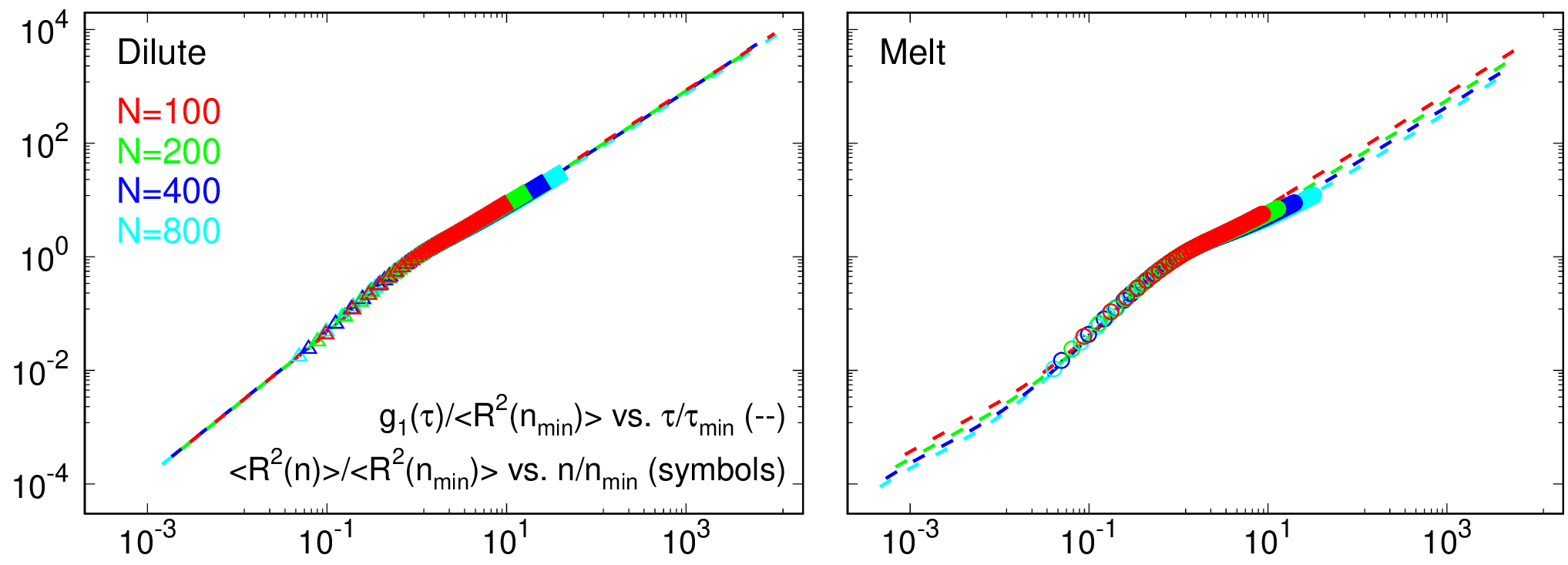} \\
\\
\\
{\rm Pe}=10 \\
\includegraphics[width=0.75\textwidth]{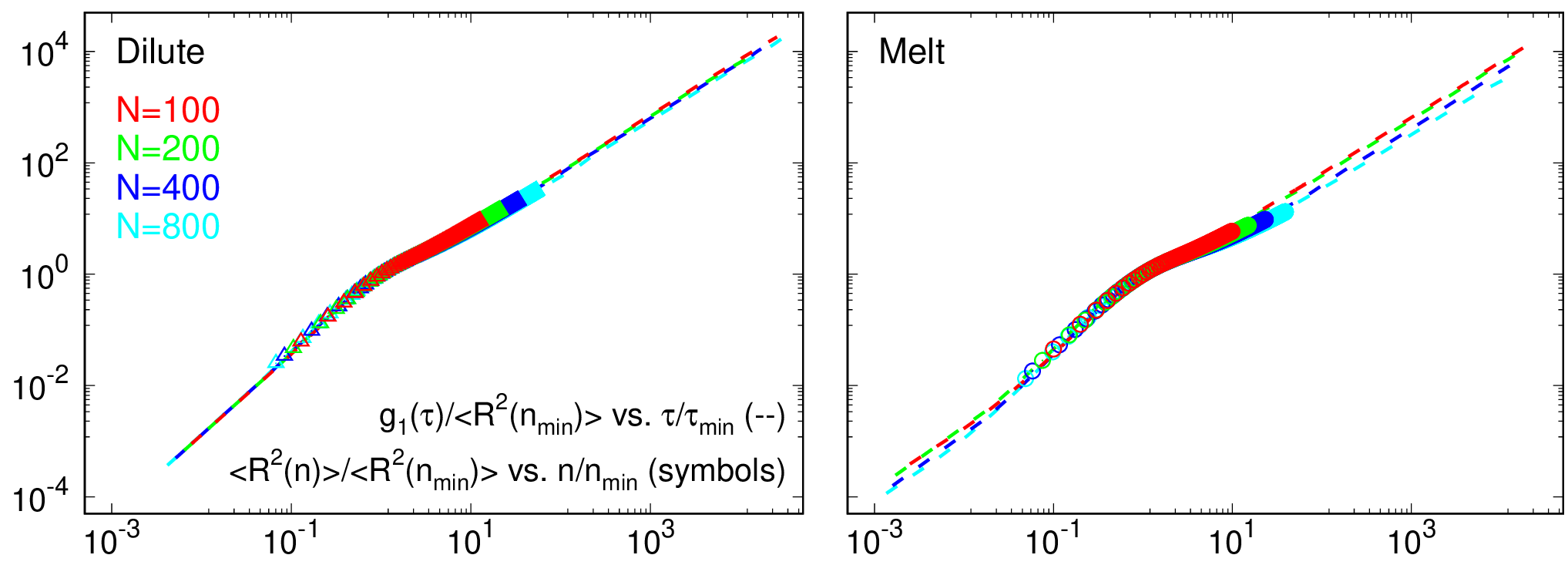}
\end{array}
$$
\caption{
Monomer mean-square displacement, $g_1$, normalized by $\langle R^2(n_{\rm min})\rangle$ as a function of rescaled time $t / \tau_{\rm min}$ (lines) in comparison to normalized mean-square internal distance, $\langle R^2(n) \rangle / \langle R^2(n_{\rm min}) \rangle$, as a function of normalized contour length separation, $n / n_{\rm min}$ (symbols, same as in the corresponding panels of Fig.~\ref{fig:R2-BCF-vs-n-scaling-Pe>0} here).
Results are for P\'eclet numbers ${\rm Pe}=1, \, 5, \, 10$ (for ${\rm Pe}=20$, see Fig.~\ref{fig:g1-scaling-Pe20} in the main paper).
}
\label{fig:g1-scaling-Pe>0}
\end{figure*}
%

%%%
\end{document}